\documentclass{mem}
\usepackage{natbib}
\usepackage{txfonts}
\usepackage{balance}
\usepackage{graphicx}
\usepackage[a4paper,breaklinks,dvipdfm]{hyperref}
\idline{NN}{1}
\begin{document}
\def\teff{$T\rm_{eff }$}
\def\kms{$\mathrm {km\,s}^{-1}$}

\title{
A super-Li rich turnoff star in NGC 6397 -- \\
the puzzle persists}

   \subtitle{}

\author{
A. Koch\inst{1} 
\and
K. Lind\inst{2}
\and
I.B. Thompson\inst{3}
\and
R.M. Rich\inst{4}
 }

  \offprints{A. Koch}

\institute{
Zentrum f\"ur Astronomie der Universit\"at Heidelberg,  Landessternwarte, K\"onigstuhl 12, 69117 Heidelberg, Germany
\and
Max-Planck-Institut f\"ur Astrophysik, Garching, Germany 
\and
Observatories of the Carnegie Institution of Washington, Pasadena, CA, USA
\and
University of California Los Angeles, Department of Physics \& Astronomy, Los Angeles, CA, USA
\email{akoch@lsw.uni-heidelberg.de}
}

\authorrunning{A. Koch et al.}

\titlerunning{A super-Li rich turnoff star in NGC 6397}

\abstract{
This presentation focuses on a recently discovered super-Li rich turnoff star in the old, metal poor 
globular cluster NGC 6397 (Koch et al. 2011, ApJL, 738, L29).  Its unusually high 
NLTE lithium abundance of A($^7$Li) = 4.21, 
the highest Li enhancement found in a Galactic GC dwarf star to date, has defied any unambiguous 
explanation through canonical enrichment channels. 
Spectra of the star show no convincing
evidence for binarity, and measured line strengths and chemical 
element abundance ratios are fully compatible with other turnoff stars in this GC, seemingly ruling 
out mass transfer from an AGB companion as origin of the high A(Li). 
A possible cause is an interaction with a red giant that has undergone cool bottom processing. 
\keywords{Stars: abundances --- stars: binaries ---  stars: Population II --- globular clusters: individual (NGC~6397) --- nuclear reactions, nucleosynthesis, abundances}
}
\maketitle{}

\section{Introduction}
The occurrence of a ``plateau'' in the Li abundance of dwarf stars over a broad range of stellar parameters has become well-established (Spite \& Spite 1982) and 
is often interpreted as due to a depletion in Li from the primordial, cosmological value in the course of stellar evolution (e.g.,  Charbonnel et al. 2005; Cyburt et al. 2008). 
As stars evolve to the red giant branch (RGB), dredge ups and mixing of Li into the hotter stellar regions, where it is easily destroyed  at a few $\times10^6$ K, 
lead to a further, strong depletion of  this element.

In this contribution, however, we shall focus on poorly understood  {\em over}abundances of Li. 
Li-rich giants are by now known to exist in the  Milky Way disk, bulge,  and halo and in  globular clusters (GCs; Kraft et al. 1999;  
Gonzalez et al. 2009; Kumar et al. 2011; 
Monaco et al. 2011; Ruchti et al. 2011, and references therein). 
In fact, statistics tells us that $\sim$1\% of all RGB stars are Li-rich, i.e., A(Li)$>$1.5 dex (Brown et al. 1989). 
On the other hand, hardly any Li-rich dwarf stars are known; the record holder until recently was a star in a young (700 Myr) open cluster, 
at an A$_{\rm LTE}$(Li) of 4.29 (Deliyannis et al. 2002).  

Here we report on a turnoff star in the metal poor GC NGC~6397, which shows an extraordinary enhancement in its surface Li abundance. This star,
first discovered and analysed by Koch et al. (2011), is amongst the most Li-rich stars known to exist in the Galaxy and in the following we will 
investigate several scenarios that could in principle cause such an overabundance. 
\section{NGC 6397}
\subsection{The cluster}
NGC~6397 is the second closest Galactic globular cluster to the observer (at R$_{\rm GC}= 6.0$ kpc; d$_{\odot}$=2.3 kpc; Harris 1996 [2010 edition]) 
and, as such, a well studied system. Numerous studies have established it as an archetypical metal-poor ([Fe/H] = $-2.1$ dex; Koch \& McWilliam 2011) 
halo object. Its stars are $\alpha$-enhanced to the canonical plateau value of [$\alpha$/Fe]$\sim$0.4 dex (e.g., Lind et al. 2011; Koch \& McWilliam 2011). 
Furthermore, it exhibits a pronounced Na-O anticorrelation (e.g., Carretta et al. 2009; Lind et al. 2011)
and shows evidence for a mild Na-Li anticorrelation (Lind et al. 2009a). 
Finally, we note the presence of significant 
trends of chemical element abundances with effective temperature and stellar evolutionary status, 
arguing for the importance of extra-mixing and diffusive processes for explaining depletions in Li and other elements
  (Korn et al. 2007; Nordlander et al. 2012 in these proceedings).  
\subsection{The data} 
We obtained spectra of  three red giants and three turnoff stars in NGC~ 6397 in July 2005 with the MIKE instrument at the 6.5-m Magellan/Clay Telescope. 
These data were presented in Koch \& McWilliam (2011) and Koch et al. (2011) and we refer the reader to these works for details on the observation, reduction, and 
analysis strategies. 
We note that those stars were chosen to have parameters close to Arcturus and the halo field turnoff star Hip 66815 
to provide a sound set of  reference stars for 
a precise,  differential abundance analyses. 
The star presented in Koch et al. (2011) and discussed in this contribution has stellar parameters (\teff, log\,$g$, $\xi$, [Fe/H]) = (6282 K, 4.1, 1.2 \kms, $-1.93$) 
and a V-band magnitude of 16.3. In particular, all three GC turnoff stars have (B$-$V) colors and magnitudes that are identical to within 0.01 mag and their 
spectroscopic \teff, log\,$g$, and $\xi$ agree to within (30 K, 0.15 dex, $<$0.2 \kms).

The serendipitous inclusion of one such chemical oddball in a sample of six targets may then seem a lucky strike in comparison with large-number surveys that 
yield $\sim$1\% of Li-rich objects. 
\section{Li in NGC 6397}
Fig.~1 shows the NLTE Li abundances we derived from our entire sample in comparison with the recent measurements of 349  
stars in NGC 6397 across a broad range of evolutionary stages by Lind et al. (2009a). 
In practice, A(Li) was determined from the equivalent widths of the 6707.7\AA~doublet and, in the case of the super-Li rich star, 
supported by the width of the 6103\AA~subordinate line. 
All values were  confirmed by spectral syntheses and corrected for NLTE following the prescriptions of 
Lind et al. (2009b; see Koch et al. 2011 for details). 
For the red giants, only upper limits could be determined. 
\begin{figure*}[t!]
\resizebox{\hsize}{!}{\includegraphics[clip=true]{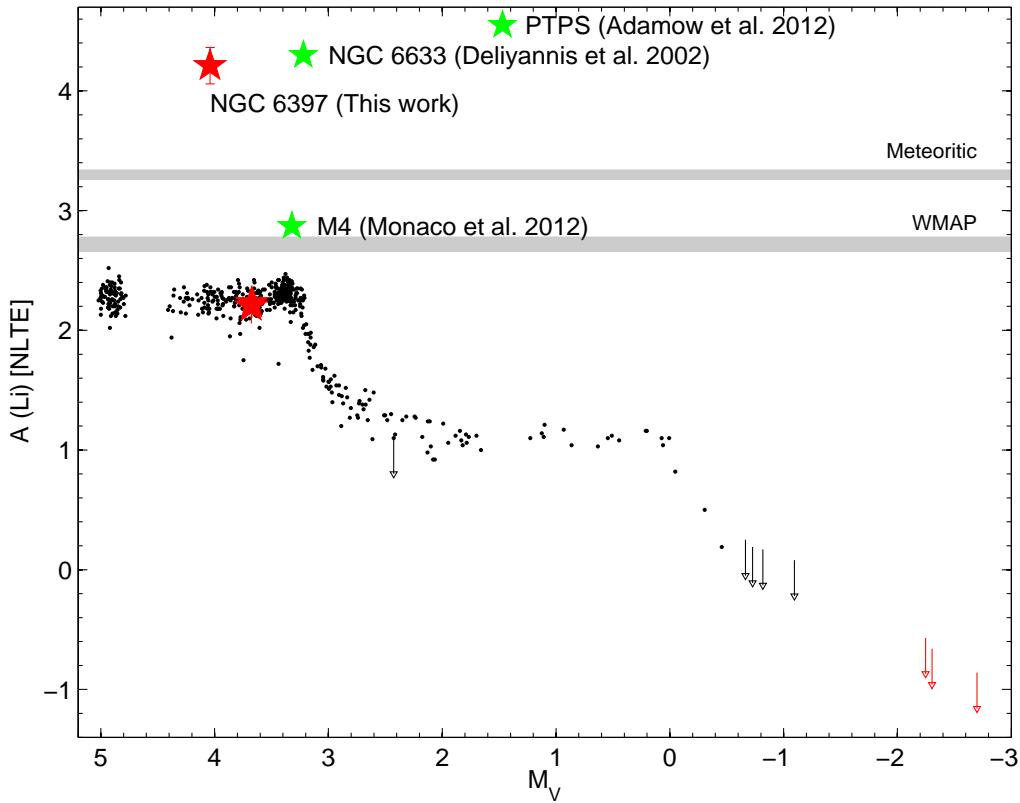}}
\caption{\footnotesize
Li abundances in NGC 6397 stars from Lind et al. (2009a; small dots); the other TO stars from Koch et al. (2011) lie on the plateau (red star symbol), 
while the brightest three RGB stars (red arrows) are heavily depleted. We highlight as
green symbols a few other remarkably over-abundant stars (Deliyannis et al. 2002; Adam\'ow et al. 2012 [these proceedings]; Monaco et al. 2012). }
\end{figure*}
\subsection{Regular Li-abundances}
The literature data follow the well established trend of the Li-plateau for
dwarf stars that transits into the depleted values on the giant branch due to the destruction of Li in the hotter interiors of the evolving stars; 
furthermore, a slight increase at the subgiant branch (at $M_V\sim3.4$ mag) is observed, in accordance with  models that 
account for diffusion and turbulence (e.g., Richard et al. 2005). 

Five out of six stars bear little surprise: the three bright giants, near the tip of the RGB  are strongly depleted in Li, at a typical A(Li)$<-0.7$ dex.
Furthermore, two of the turnoff stars lie square on the plateau-value at A(Li)=2.2 dex. 
\subsection{Li-overenhancement}
One of the turnoff stars, \#1657 (Kaluzny 1997) $\equiv$ 2MASS J17410651-5343290, shows an extraordinarily high Li abundance of A$_{\rm NLTE}$(Li)=4.21 
dex. This accounts for continuum veiling by a contaminating foreground star (see Sect. 4.4). This is the largest value for A(Li) found in a GC dwarf star 
to date.  

Figure~1 also highlights the other few super-Li rich stars in the literature that are found in a variety of environments. 
These are a dwarf in the young (700 Myr) open cluster NGC~6633 (Deliyannis et al. 2002); a dwarf in the GC M4, which shows the primordial Li-value 
(Monaco et al. 2012); and one highly enriched disk  giant from the Pennsylvania-Toru\'n Planet Search (PTPS; Adam\'ow et al. 2012; these proceedings).
Note that, while the data by Monaco et al. (2012) account for NLTE, no such corrections have been applied to the 
stars reported by Deliyannis et al. (2002) and Adam\'ow et al. (2012). These are likely of the order of $-$0.2 dex (Lind et al. 2009b). 
\section{Possible reasons for an enhancement}
Such high levels of Li-enhancements as found in \#1657 are not easily understood in terms of simple 
evolutionary mechanisms and a variety of environments can be evoked as sources of the extra lithium. 
 \subsection{Planetary ingestion}
While the accretion of asteroidal material (e.g., Jura et al. 2012) or an 
engulfment of planets or brown dwarfs can in principle enhance the surface Li abundance and/or trigger 
fresh nucleosynthetic processes (Siess \& Livio 1999), the amount required to elevate the Li abundance in our object 
would probably require an unrealistically massive object to be accreted.
Furthermore, the occurrence of such substellar bodies is in very unlikely in metal poor environments such as NGC 6397. 
Moreover, there is no evidence of circumstellar material visible in the Na D lines. 
A clear-cut requirement for  this scenario to work for \#1657 
would be the detection of a large Beryllium enhancement in this star  (Ashwell et al. 2005). 
This is not seen. %
 \subsection{Type II Supernovae}
In principle, core-collapse Supernovae of type II in the mass range around 30 M$_{\odot}$
are able to manufacture sufficient large amounts of Li in the $\nu$-process 
 (Woosley \& Weaver 1995). On the other hand,  such massive SNe II would also 
 yield elevated abundances of the hydrostatic ($\alpha$-) elements; this is incompatible with the regular Mg-abundance found in this star (Koch et al. 2011), which 
  is in agreement with the values found in the rest of the sample of Koch \& McWilliam (2011).
  \subsection{Diffusion / Radiative acceleration}

 Diffusive models (e.g., Richer \& Michaud 1993; Richard et al. 2005) predict 
the surface convection zone of dwarfs to be enriched by radiative outward acceleration of Li from deeper regions. 
However, this process is most efficient in a  narrow temperature window of 6900 -- 7100 K (e.g., Deliyannis et al. 2002) and, at 
 \teff=6280$\pm$250 K, our Li-rich star is too cool for this acceleration to occur. 
  \subsection{Binary transfer}
Is it possible that the Li visible in our star has been produced externally and has been transferred onto its surface?
\subsubsection{Asymptotic Giant Branch companion}
The standard mechanism to produce Li has been conceived by Cameron \& Fowler (1971)  to take place in 
the outer, convected layers of Asymptotic Giant Branch (AGB) stars. Likewise,  Li-enrichment in the wind of super-AGB stars 
 (i.e., massive AGBs of $\sim$6--8 M$_{\odot}$)  that experienced hot bottom processing can be a possible mechanism (Ventura \& D'Antona 2011).  

Could the Li in the GC turnoff star then originate in a companion that has already evolved through the AGB phase and faded since? 
In fact, there is a clear contaminant visible in the spectrum at a radial velocity of $-102$ \kms~(Fig.2, top), but at the large velocity difference of 120 \kms~to the GC star 
it is highly improbable to  be in a physical binary. Furthermore, the spectra do not show any 
other evidence of a faint companion and there is no significant velocity variation over the 6+ years of our study (Fig.~2, bottom).  
Finally, we note that, although  photometric data exist for  this star, e.g.,  from the  HST, 
these are not sufficient to assess variability (V. Nascimbeni, priv. comm.) and we 
cannot conclude on the binarity on this object from this point of view.  
\begin{figure*}[t!]
\begin{center}
\resizebox{0.7\hsize}{!}{\includegraphics[clip=true]{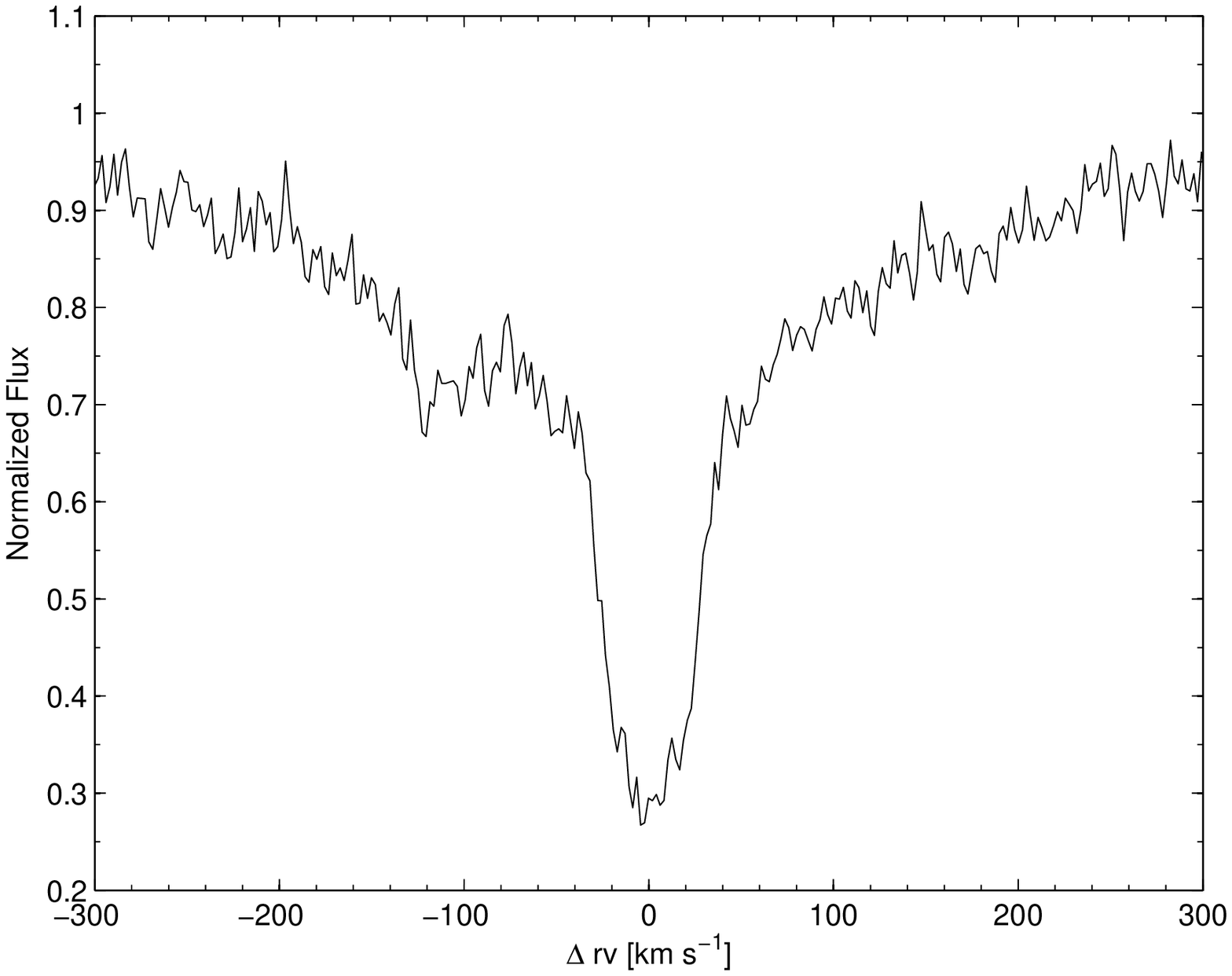}}
\resizebox{0.78\hsize}{!}{\includegraphics[clip=true]{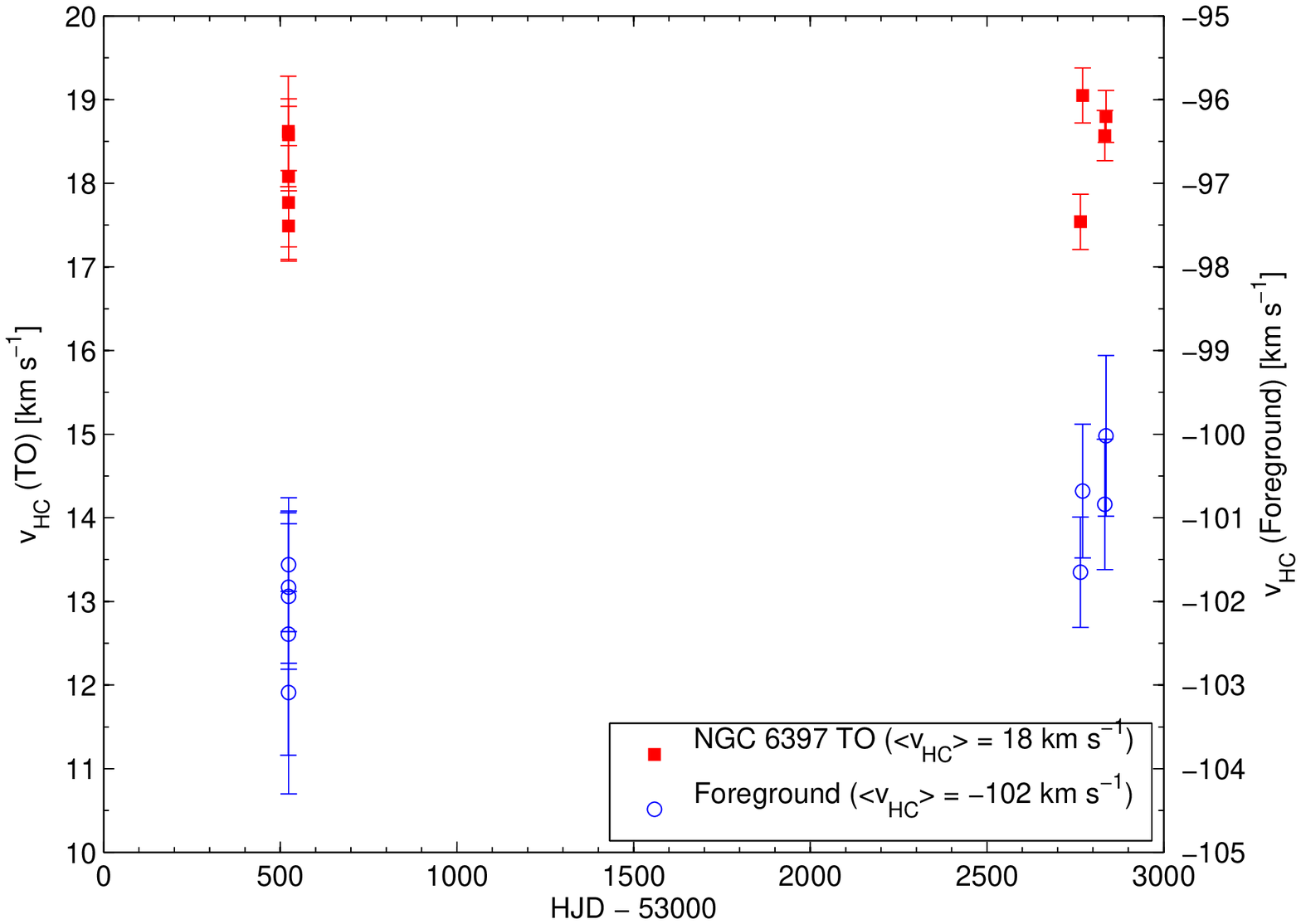}}
\end{center}
\caption{\footnotesize
{\em Top panel:} H$\alpha$ profile of the turnoff star; note the contaminating foreground star at $\Delta$\,rv=$-120$ \kms. 
This object does not contribute more than 16\% to the continuum flux, which was accounted for in all our abundance measurements.  None of 
its spectral lines overlap with any of the features discussed here and in Koch et al. (2011). 
{\em Bottom panel:} Radial velocity data for the GC star (red) and the foreground contaminant (blue). The latter points were offset vertically for clarity. 
}
\end{figure*}

If the Li in question had been produced through hot bottom burning  one might expect to see accompanying enrichment patterns in the C,N,O abundances, which are 
inaccessible from the current spectra.  Such AGB nucleosynthesis 
would also yield enhanced s-process element abundances (although the s-process elements would  not be dredged up in the 
{\em super-}AGB stars), but as Fig.~3 indicates, there is no such evidence in the stars.
\begin{figure}[t!]
\begin{center}
\resizebox{\hsize}{!}{\includegraphics[clip=true]{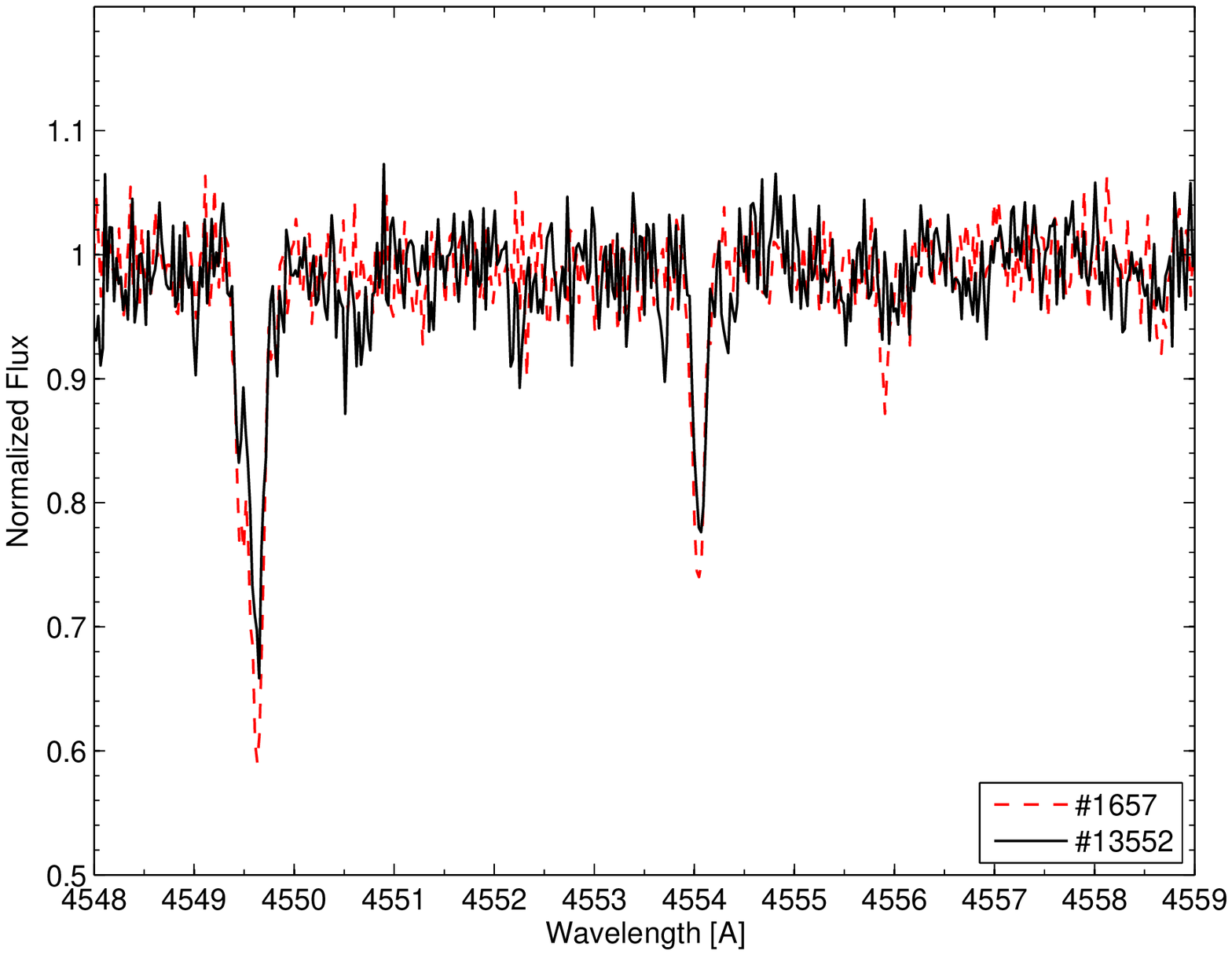}}
\end{center}
\caption{\footnotesize
Spectral region around the Ba\,{\sc ii} line at 4554\AA. The dashed line is the Li-rich turnoff star \#1657 (Koch et al. 2011), while the solid line is the regular 
turnoff star \#13552 from Koch \& McWilliam (2011). The difference in line strength is solely due to the continuum veiling from the non-associated foreground contaminant. 
}
\end{figure}

Furthermore, we find a low [Na/Fe] abundance ratio of $\sim -0.59$ dex, which designates this star as an object from the GC's primordial generation 
(Carretta et al. 2009), such as one of the other turnoff stars, while the remainder of the sample are second generation, i.e., high-Na, low-O stars. 
Unfortunately, O is not measurable in the spectra of the Li-rich turnoff star.
\subsubsection{Red Giant Branch companion}
Li synthesis via the Be-transport mechanism (Cameron \& Fowler 1971) can also proceed 
 in a H-shell fusion zone in low-mass giants, as soon as  extra deep envelope mixing becomes viable. 
 This  self-enrichment process (``cool bottom processing'' [CBP]; Sackman \& Boothroyd 1999) 
  is particularly efficient in metal poor GC stars, since they have hotter CNO-burning shells. 
In our star, mass transfer from  a red giant companion that underwent CBP
is an attractive possibility, as this process does not incur enhancements of the $s$-process elements, while it retains 
the regular abundance patterns emerging from canonical RGB nucleosynthesis. 
 
 However,  Li is destroyed by proton capture over very short periods  on the RGB, at typical depletion time scales of a few $\times 10^4$ yr (e.g., Kraft et al. 1999; and 
references therein), as the freshly synthesized Li is mixed back deeply into the hotter layers.
Thus, while the chemical signatures in our star may argue in favor of an external enrichment by CBP-processed material, a fortunate timing of the mass transfer  
process during the common, yet transient, phase of Li-enhancement on the RGB must be realized. 

Interestingly, the binary fraction of NGC 6397 is estimated to be only 1-5\% (Richer 2008), i.e. similar to the fraction of Li-rich red giants (although the latter 
have not been discovered in the cluster yet). This leaves only a very small probability 
that the extra Li has been produced prior to mass transfer by CBP in a red giant. 
In this context of binaries 
we note the hypothesis that Li can be produced by spallation on the surfaces of late-type stars in binary pairs with a neutron star or black hole companion 
(Martin et al. 1994; Fujimoto et al. 2008), but we will not 
discuss the implications any further in this place.

\section{Discussion and outlook}
In this contribution we have summarized our recent findings regarding the super-Li turnoff star in the metal poor GC NGC~6397 (as reported in Koch et al. 2011). 
None of the standard scenarios generally evoked to explain Li-overabundances in excess of the primordial value, i.e., A(Li)$\ga$2.7 dex
appears to work in this object. All the abundance patterns we detect are 
very similar to those found in the other stars of our sample in the same GC, seemingly ruling out
planetary ingestion, radiative acceleration, SNe II enrichment, or the pollution from a nearby AGB companion. 
While the latter AGB contamination seems incompatible with the low Na-abundance we find, transfer from a former RGB star is not ruled out. However, this requires a delicate timing of all processes involved and, furthermore, our radial velocity data is inconclusive as to the binary nature of this object. 

The best way to test these scenarios against each other is then  to measure  CNO and Beryllium abundances -- these can give clues as to the nature 
of any potential polluter and help to assess the significance of planetary accretion. 
Another scenario that has been mentioned is the production of light elements due to  short-range shock-wave induced spallation through the 
ejecta of SNe II (e.g., Fields et al. 1996). The chemical imprints of this mechanism on our star, however, requires more study. 

The final, lingering question is of course:  why aren't there more dwarfs known with such high Li-enhancements?  
Recent studies have emphasized the potential of uncovering Li-rich giants and a few Li-rich dwarfs are counted (see Sect.~1), 
although a standardized explanation for their enrichment is yet to be found.
 Future survey missions such as  4MOST (de Jong et al. 2012) 
are well suited to detect such chemical oddballs in larger numbers and to characterize to first order their chemical peculiarities. 
\begin{acknowledgements}
AK thanks the Deutsche Forschungsgemeinschaft for funding from  Emmy-Noether grant  Ko 4161/1. 
\end{acknowledgements}

\bibliographystyle{aa}

\end{document}